# A Comparative Study of Algorithms for Intelligent Traffic Signal Control

Hrishit Chaudhuri[*,1], Vibha Masti[*,1], Vishruth Veerendranath[*,1], and S Natarajan[1]

[1]Department of Computer Science, PES University, Bengaluru, India
[1]`hrishitchaudhuri@gmail.com`, {`vibha, vishruth`}`@pesu.pes.edu`,
`natarajan@pes.edu`

**Abstract.** In this paper, methods have been explored to effectively optimise traffic signal control to minimise waiting times and queue lengths, thereby increasing traffic flow. The traffic intersection was first defined as a Markov Decision Process, and a state representation, actions and rewards were chosen. Simulation of Urban MObility (SUMO) was used to simulate an intersection and then compare a Round Robin Scheduler, a Feedback Control mechanism and two Reinforcement Learning techniques - Deep Q Network (DQN) and Advantage Actor-Critic (A2C), as the policy for the traffic signal in the simulation under different scenarios. Finally, the methods were tested on a simulation of a real-world intersection in Bengaluru, India.

## 1     Introduction

The objective of this paper is to provide a comparative analysis between four policies used to manage queue build-up in traffic signals. This work aims to identify the relative strengths and weaknesses of the current state-of-the-art Reinforcement Learning (henceforth, RL) approaches to this problem, against those of simpler non-RL based approaches. This work is necessitated to determine if the computational intensity of RL algorithms is viable and justified. The two RL techniques chosen are the Deep Q-Network (DQN) and the Advantage Actor-Critic (A2C) techniques, compared against a Round Robin (RR) scheduler, and a Feedback Control Mechanism (FCM). All traffic data used in these simulations have been randomly generated, according to a Weibull generation. Simulations have been performed using Eclipse's traffic simulator, Simulation of Urban MObility (SUMO).

---

[*] These authors contributed equally to this work

## 2    Related Work

The first course of action was to identify the possible formalisms that could be adopted towards the problem. This is a well-known issue in tackling the signal-timing problem, and individual approaches taken by authors include Wey's Mixed-Integer Linear Programming (MILP) model [1], as well as the Tournament-Based Elimination Pairing system treatment utilized by Eriskin et al. [2]. However, the intuitive notion of each traffic signal behaving like a selfish agent intending to minimize the average waiting time of its lane allows us to formalize the intersection as a stochastic repeated game, an assumption that leads to the natural formulation of the problem as a Markov Decision Process (henceforth, MDP).

Existing research tends to differ on the fundamental definition of the MDP. In particular, authors have not settled on a universal definition for the state, action, probability, or reward spaces that comprise the MDP. As an example, Abdoos et al. [3] describe the environment in terms of the queue length (an approach also used by this project), whereas another popular description involves visualising the environment in terms of the average waiting time [4]. Upon settling on a single model for experimentation, the question of choosing an appropriate learning algorithm becomes crucial. The simplest way to approach this is to use tabular methods such as Q-learning over a discretized state space, which employs a simple search-for-maximum-reward mechanism over each epoch.

Signal-timing algorithms have traditionally been divided into *fixed-time control*, where the signal-timing plan is deduced a priori, without any feedback from the relevant environment, and the availability of real-time data does not affect the mechanism in any way, and *actuated methods*, where the algorithm also deduces a posteriori, with a set of rules provided whose violations can lead to triggering an update in the working of the algorithm.

Actuated methods have also been studied in the context of wireless sensor networks. Pandit et. al. [15] discussed the use of the Oldest Arrival First (OAF) algorithm, a derivative of the Oldest Job First (OJF) scheduling algorithm, and the usage of vehicular ad-hoc networks (VANET) to implement adaptive traffic signal control. More recently, work by Hilmani et al. [14] has explored the use of wireless sensor networks to manage traffic signals in detail.

RL approaches have also been examined [5-6, 13], with two crucial algorithms being benchmarked against the efficacies of a fixed-time control and an actuated mechanism. The first of these is a deep Q-network, based on the work by Vidali et al. [7], where a deep neural network is trained to learn the state-action function based on the generic Q-learning equation. The second is an Actor-Critic mechanism: The Advantage Actor-Critic (A2C), which stems from its asynchronous variant, Asynchronous Advantage Actor-Critic A3C [8].

Several methods for obtaining traffic data have also been explored, with three main approaches looked at: real-time traffic data from surveillance cameras [11], real-world data obtained from Google Maps and road cameras [12] and synthetic data generated according to a Weibull distribution [7]. The authors of this paper have used synthetic data generated according to a Weibull distribution for the experiments.

## 3    Experimental Setting

All the simulations involved in the undertaking of this experiment have been executed using the Eclipse Foundation's Simulation of Urban MObility (SUMO) traffic simulator [9]. Interfacing with the Python RL scripts has been performed using the Traffic Control Interface (TraCI), and data has been generated according to SUMO's internal data dumps.

The problem as defined by Vidali et al. [7, pg.44] is: "*given the state of the intersection, what is the traffic light phase that the agent should choose, selected from a fixed set of predefined actions, in order to maximize the reward and consequently optimize the traffic efficiency of the intersection.*"

Here, the state of the intersection defines the state space of the MDP, with varying definitions according to the algorithm chosen. The action space consists of choosing a particular arm of the intersection to signal green at any time. The reward function accordingly varies from algorithm to algorithm. The intersection used for experimentation is a four-way single lane intersection with a single traffic light controller in the middle as shown in Fig. 1. The length of the road in each direction is fixed at 242.8 m. The state representation for the intersection (environment) is chosen suitably based on the signal-timing algorithm employed.

The action set (1) is the set of possible actions that the agent can take from any given state $s_t$ of the environment. For our RL approaches, the action set (1) represents the possible green phases that the agent – the traffic signal – can activate.

$$A = \{S, E, N, W\} \tag{1}$$

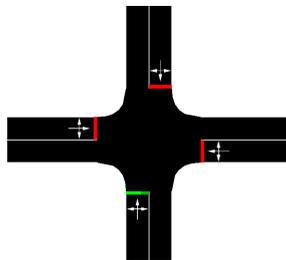

**Fig. 1.** Four-way single lane intersection (environment)

Every action of set (1) belonging to *A* represents the incoming lane (arm) for which the green phase of the traffic light is activated. For instance, the action *E* represents the agent activating the green phase for the east (E2T) lane and deactivating the green phase for the remaining three incoming lanes – north (N2T), south (S2T), and west (W2T).

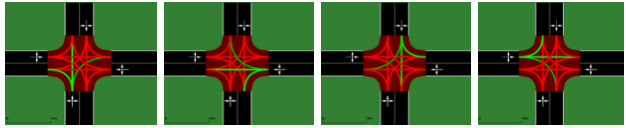

**Fig. 2.** The four green phase actions (*S, E, N, W*)

It must be noted that when the agent takes an action $a_t$ and it differs from the previously taken action $a_{t-1}$, an intermediate action of yellow phase activation must be taken. For instance, if the previously taken action was E and the next action to be taken is N then an intermediate action $E_{yellow}$ must be taken where the yellow phase for the incoming east lane (E2T) is activated.

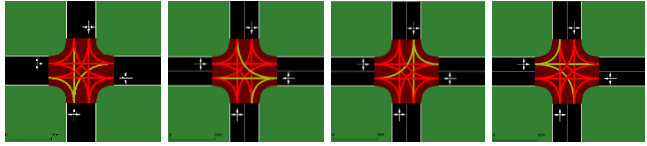

**Fig. 3.** The four intermediate yellow phase actions

Traffic route files used for testing and training are generated with vehicular density per lane plotted according to a Weibull distribution, to ease loading over time. Three different traffic scenarios are tested for each signal scheduling algorithm, with the traffic conditions under each instance referred to by a specific SCENario (SCEN) number, as done by Vidali et al. [7]. In the first case, traffic is distributed uniformly between all arms (SCEN-1). In the second, the density of traffic on the N2T-S2T arms is heavier – 90% of the vehicles travel on the N2T-S2T arms – (SCEN-2). Finally, a heavier traffic density is placed on the E2T-W2T arms (SCEN-3). A summary is shown in Table 1.

In the case of our Feedback Control approach, the action space is defined to be a range of numbers from min_time to max_time. Here, the order in which green phases are activated is fixed and only the green signal timing varies based on traffic density. In this approach, an intermediate yellow phase action is taken after every green phase action (in other words, 50% of the actions are yellow phase actions).

### 3.1 Round Robin Scheduling

In this method, each agent is assigned a fixed quantum of time during which its corresponding arm is signalled green. This quantum remains invariant under all traffic conditions, including empty arms. Each quantum consists of 30 timesteps of signalling green, followed by 3 timesteps of signalling yellow. No distinction is made between right-turning, straight-going, and left-turning traffic. The experiment is repeated thrice, with each of the traffic scenarios as described above (SCEN-1, SCEN-2, and SCEN-3).

**Table 1.** Traffic probability for a single car

| Direction | Source | Destination | Probability in SCEN-1 | Probability in SCEN-2 | Probability in SCEN-3 |
|---|---|---|---|---|---|
| Straight | North | South | 0.1875 | 0.3375 | 0.0375 |
| | South | North | 0.1875 | 0.3375 | 0.0375 |
| | East | West | 0.1875 | 0.0375 | 0.3375 |
| | West | East | 0.1875 | 0.0375 | 0.3375 |
| Turn (Left or Right) | North | East | 0.03125 | 0.05625 | 0.00625 |
| | North | West | 0.03125 | 0.05625 | 0.00625 |
| | South | West | 0.03125 | 0.05625 | 0.00625 |
| | South | East | 0.03125 | 0.05625 | 0.00625 |
| | East | North | 0.03125 | 0.00625 | 0.05625 |
| | East | South | 0.03125 | 0.00625 | 0.05625 |
| | West | South | 0.03125 | 0.00625 | 0.05625 |
| | West | North | 0.03125 | 0.00625 | 0.05625 |

Round Robin (RR) was explored and compared against other methods in this paper since RR is the most prevalent traffic signalling algorithm used in most traffic intersections around the world. Using RR as a baseline enables us to effectively evaluate the relative performance improvements offered by other methods.

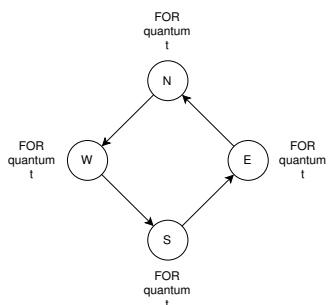

**Fig. 4.** Round Robin Scheduling for time quantum t

### 3.2 Feedback Control Mechanism

This is a slight modification of RR. The primary idea of this approach is to enable the signal to perform some basic "learning" by reading inputs from the environment and updating its actions based on those inputs. For this reason, the action space is selected as an array of discrete time intervals which the traffic agent can select from. The action represents the length of time the agent signals green.

This model uses a game-playing strategy, where each agent is given a 'turn' in RR fashion, similar to players in most board games, such as Monopoly. For this reason, this mechanism will be referred to as MONOPOLY, henceforth. The MONOPOLY reward function searches over all actions in the agent's reward space and calculates the reward for each possible action. The reward function is defined in (2) where $s_t$ represents the state of the corresponding lane at time $t$. Here, the state

$$R(s_t, a) = | s_t - v_t\, a | \qquad (2)$$

of the lane only refers to the length of the queue at time $t$. $v_t$ refers to the average speed of the cars in the junction at the last timestep, and $a$ is the action currently being compared. The action generating the maximum value of $R$ implies that the signal neither overshoots nor undershoots but stays green for exactly the time required to clear out the queues.

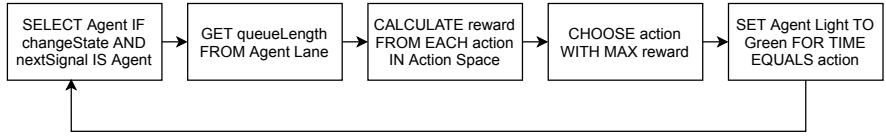

**Fig. 5.** Feedback Control Loop

### 3.3 Deep Q-Learning

In a reinforcement learning model, an agent is placed in an environment and is made to observe its state $s_t$. The agent then proceeds to choose an action $a_t$ from a given set of actions according to a Markov Decision Process (MDP), transitioning the environment from the current state $s_t$ to the next state $s_{t+1}$. Upon making the transition, the agent receives a reward $r_{t+1}$ based on the previous action $a_t$. The goal of the agent is to choose a sequence of actions in the environment that maximizes the reward.

Q-learning is a reinforcement learning technique where at every step the value function (known as the Q-function) is updated following an equation (typically the Bellman Equation). In regular Q-learning, a table of Q-values $Q(s, a)$ is maintained to estimate the optimal Q-function $Q^*(s, a)$ and the entries of the table are updated as the agent learns. As the state space gets larger, maintaining a table becomes infeasible. Deep Q-learning makes use of a deep neural network to estimate the value of the optimal Q-function $Q^*(s, a)$. In this paper, the Q-function used for Deep Q-learning (3) is the one proposed by Vidali et al. [7].

$$Q(s_t, a_t) = r_{t+1} + \gamma \cdot \max_A Q'(s_{t+1}, a_{t+1}) \qquad (3)$$

$Q'(s_{t+1}, a_{t+1})$ represents the Q-value of taking the action $a_{t+1}$ in the next state $s_{t+1}$, the term $r_{t+1}$ represents the reward obtained by taking the action $a_t$ from state $s_t$, the term $Q(s_t, a_t)$ represents the Q-value of the action $a_t$ taken from the state $s_t$. The term $max_A$ refers to the action $a_t$ out of all possible actions taken in state $s_{t+1}$ that maximizes the value of $Q'(s_{t+1}, a_{t+1})$. The value γ represents a discount factor between 0 and 1 that diminishes the weightage of future rewards and emphasizes the immediate reward. The value α represents a learning rate between 0 and 1 that updates the current Q-value with a discounted learned value. As the agent learns and updates the Q-table, it acts as a better estimate for the optimal Q-function $Q^*(s, a)$.

The state describes a snapshot of the environment (in this case, the four-way intersection serves as the environment) at a given timestep $t$ and is represented as $s_t$. The state should give an accurate description of the number of cars in each lane at a given timestamp $t$. The state representation used is similar to the one described by Vidali et al. [7]. Each incoming road is divided into cells of varying size, where the cells are smaller and more concentrated closer to the traffic light and get larger as the distance from the traffic light increases.

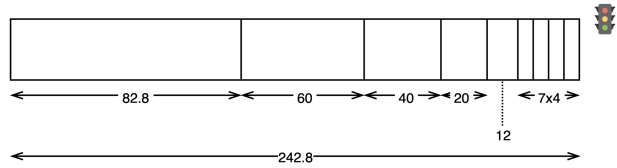

**Fig. 6.** State Representation (9 cells)

Each lane consists of 9 cells leading to a total of 36 cells in the environment of 4 incoming lanes. The state can be represented as a boolean vector of length 36 where the value of each cell in the vector is 1 if a vehicle is present in the cell and 0 if the cell is empty. Using this state representation, a total of $2^{36}$ states exist in the state space. Computationally, having a boolean state-space vector is much simpler than simply having a state vector of length 4 where each entry is the number of vehicles in a given lane.

The model used to train the agent for Deep-Q learning is a neural network of 1 input layer, 5 hidden dense layers, and 1 output layer (Fig. 7). The input to the network is a vector of length 36 and the output from the network is a vector of length 4, representing the 4 possible actions that can be taken at every stage. The green phase duration for every arm is a multiple of 5 timesteps. At every step, the agent is trained to decide which arm of the signal to activate the green phase for.

The agent is trained for 100 episodes, alternating between the three traffic scenarios, and in each episode, the experience replay is sampled from 800 times. The value of γ is set at 0.75 and a learning rate of 0.001 is used.

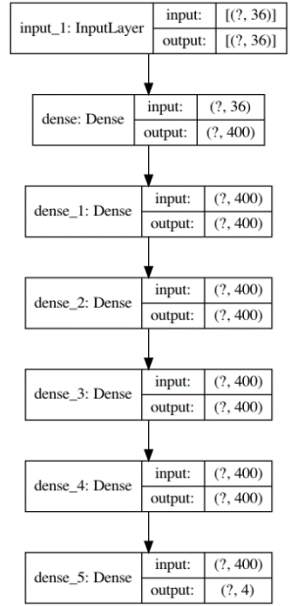

**Fig. 7.** Neural Network (DQN) Architecture

**3.4 Advantage Actor-Critic (A2C)**

The Advantage Actor-Critic mechanism belongs to a family of RL algorithms collectively known as Actor-Critic algorithms. These algorithms offer a method to bifurcate the processes of learning and evaluating strategies or actions and their relative effects on the environment. Two separate processes, termed the 'actor' and 'critic' processes, are employed in this case. The actor method is responsible for *choosing* a particular strategy. The critic method then evaluates the effects of that action (by calculating something akin to a Q-value in the DQN approach), measures the regret attached to that action, and sends this feedback to the actor process.

The Advantage function in A2C is a modification to this actor-critic architecture making use of Temporal Difference Learning (TDL). In TDL, the actor now tries to *predict* rewards. It assumes that the current reward also includes some discounted future reward that it may achieve by moving into a particular *next state*. The critic's work is to calculate this next state reward valuation. An error, termed the temporal difference error, is now introduced as a parameter: it is the difference between the predicted value of all future rewards from the current state and the current value of the state, i.e., the extra reward gained by taking that particular action, over the average reward that would be gained from the other actions. This is the *advantage* of the state. The actor can now learn by measuring its advantage: if the advantage is greater than zero, it implies that the action leading to this advantage is good for the agent, and therefore the agent should continue choosing this action.

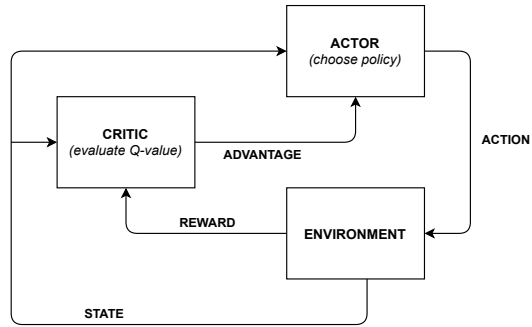

**Fig. 8.** A2C architecture

The Actor-Critic method also provides an opportunity to incorporate parallelism into Deep Reinforcement Learning as described in Minh et al. [8]. This is achieved by using multiple agents that learn on their independent instances of the environment in parallel. An agent-environment pair is referred to as a worker. Since the environment for each agent is independent, the experiences of the agents differ, enabling the model to learn on a larger set of states in less time. This decreases training time significantly, also enabling the training of Deep Q-Learning networks on a CPU rather than a GPU.

A2C synchronously updates the global neural network, by averaging the weights of all the workers, while in A3C the workers asynchronously update the network by themselves. The agent is trained on the three scenarios described earlier using SUMO-RL [10] and the three scenarios are tested on it.

## 4  Results

The results of testing the different traffic light agents (RR, MONOPOLY, DQN, A2C) in the three traffic scenarios (SCEN-1, SCEN-2, and SCEN-3) for a time period of 5400 timesteps are compared. The main parameters of interest are the peak total queueing length of vehicles waiting in all four incoming arms of the intersection, the time spent above half queue length, and the similarity in the graphs of all traffic scenarios.

### 4.1  Round Robin Scheduling

The agent is made to perform in the three traffic scenarios and the total queueing lengths are observed in Fig. 9 (queue length vs timestep). The peak total queue length in SCEN-1 is around 88 vehicles, whereas for SCEN-2 and SCEN-3 the peak total queue length is around 57 vehicles. The agent performs well in uniform traffic scenarios (SCEN-1) and the queue length is above half only 33% of the time, but in non-uniform scenarios (SCEN-2 and SCEN-3), it is above half ~80% of the time. The shapes of the uniform and non-uniform graphs are not similar.

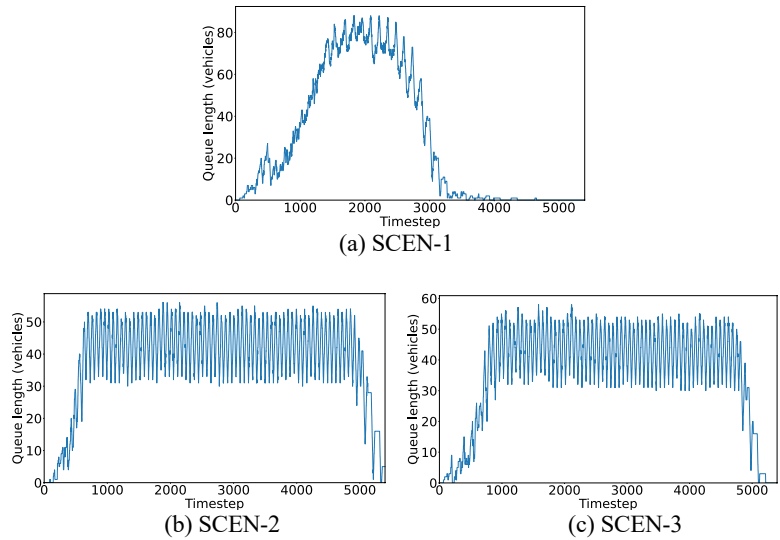

(a) SCEN-1

(b) SCEN-2

(c) SCEN-3

**Fig. 9.** Total queue lengths (y-axis) vs timestep (x-axis) using Round Robin (RR)

### 4.2 Feedback Control Mechanism

MONOPOLY was tested against SCEN-1, SCEN-2, and SCEN-3 traffic density conditions, with the plots in Fig. 10 (queue length vs timestep).

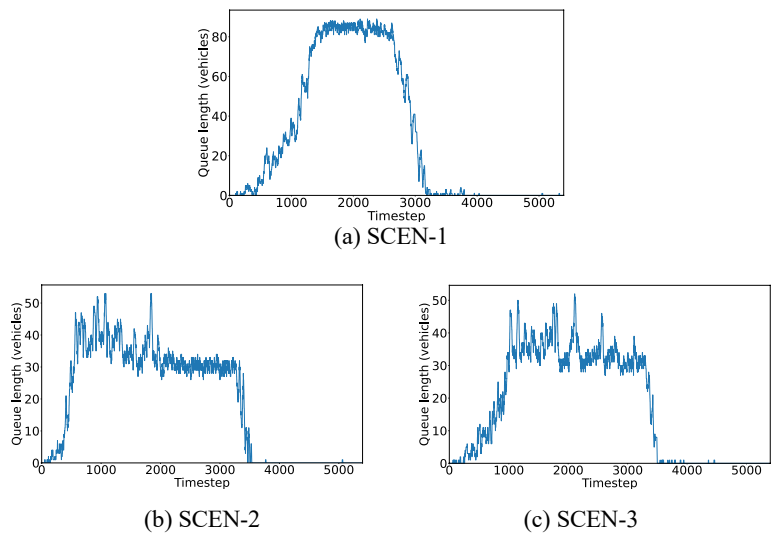

(a) SCEN-1

(b) SCEN-2

(c) SCEN-3

**Fig. 10.** Total queue lengths (y-axis) vs timestep (x-axis) using Feedback Control

The performance of MONOPOLY in uniform traffic scenarios (SCEN-1) displays a similarity in behaviour to RR. However, under non-uniform traffic densities, the agent performs significantly better than the RR agent. The time spent with a queue length above half is 50% because green signalling time is not wasted unnecessarily on empty lanes. The shapes of the graphs in uniform and non-uniform traffic are quite similar. The peak total queue length under uniform traffic is around 89 vehicles and for non-uniform traffic, it is around 53 vehicles.

A significant disadvantage of MONOPOLY is the fact that under high traffic loads, the agents act selfishly, with each agent signalling green for the maximum permissible time, causing the mechanism to revert to RR.

### 4.3   Deep Q-Networks (DQN)

The DQN agent was tested against SCEN-1, SCEN-2, and SCEN-3 traffic scenarios, and the corresponding queuing lengths are shown in Fig. 11 (queue length vs timestep). The results of testing the DQN agent on the uniform traffic agent show no major improvement over RR and MONOPOLY. However, its performance in non-uniform traffic densities is significantly improved. The queues of traffic in all four arms of the intersection eventually clear out to completion or near completion by around 4000 timesteps, and the graphs of all three traffic scenarios are quite similarly shaped. The peak total queue length under uniform traffic is around 90 vehicles and for non-uniform traffic, it is around 60 vehicles.

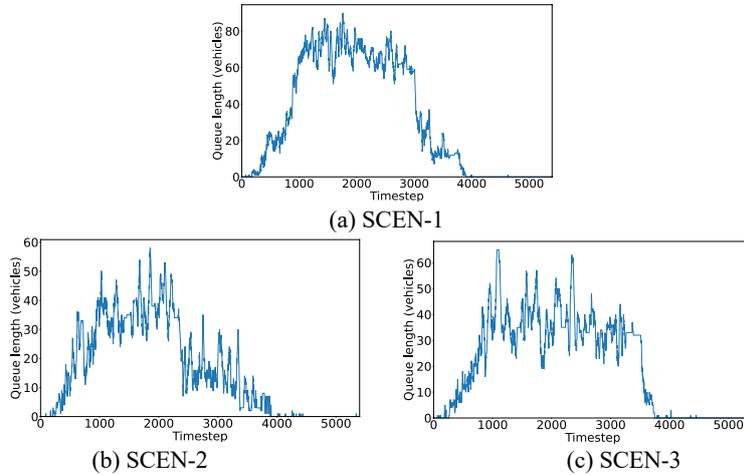

**Fig. 11.** Total queue lengths (y-axis) vs timestep (x-axis) using DQN

An important advantage of the DQN agent is in the symmetry of its performance in the three traffic scenarios. The uniform and non-uniform plots are close to indistinguishable, with around 30-40% of the time spent with a queue length above half. However, the training process takes a long time and as the networks get more complicated, the training time is expected to increase drastically.

### 4.4 Advantage Actor-Critic (A2C)

The A2C agent was tested against the same three traffic scenarios. The plots of queue length over time (Fig. 12) were observed to be similar to DQN in all three scenarios when the number of processes (workers) was set to 1 since the underlying mechanism of A2C is a Multilayer Perceptron (MLP) Neural Network. Though the queue lengths were the same, the training time was significantly lesser than DQN.

A performance improvement was observed when the number of processes (N-CPU) was increased. The performance measure chosen for the comparison was the total wait time, instead of the queue lengths. The results of the comparison between the different number of processes are shown in Fig. 13-15 (waiting time vs timestep and waiting time vs no. of processes).

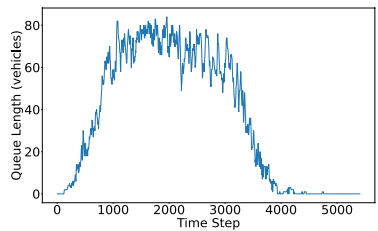

(a) SCEN-1

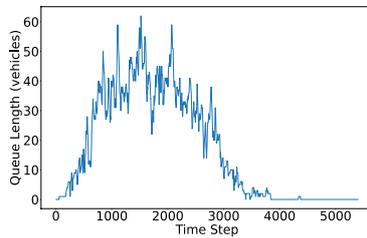

(b) SCEN-2

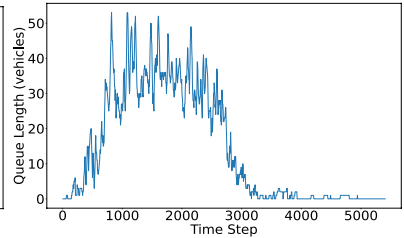

(c) SCEN-3

**Fig. 12.** Total queue lengths (y-axis) vs timestep (x-axis) using A2C

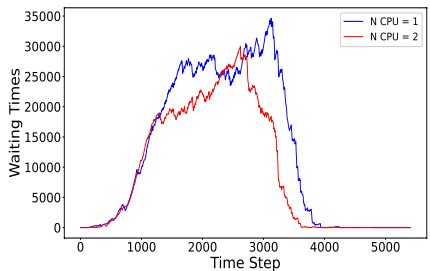

**Fig. 13.** N-CPU = 1 vs N-CPU = 2

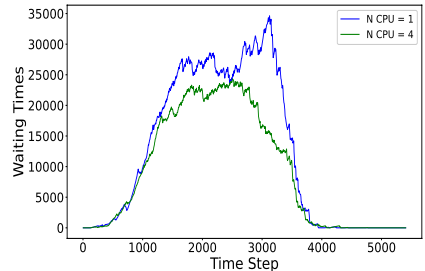

**Fig. 14.** N-CPU = 1 vs N-CPU = 4

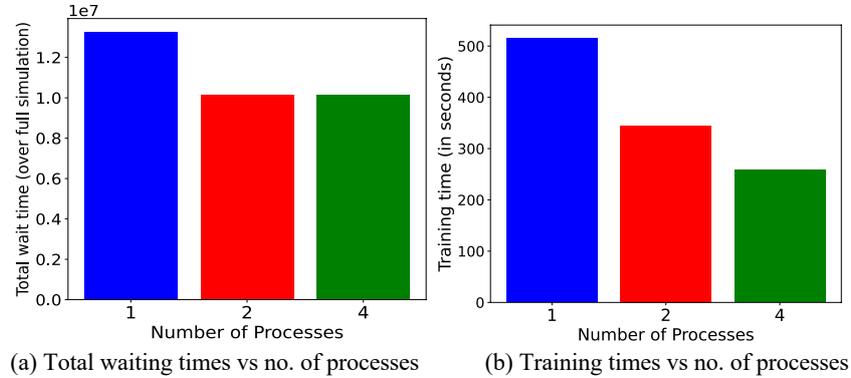

(a) Total waiting times vs no. of processes  (b) Training times vs no. of processes

**Fig. 15.** Comparison of N-CPU = 1, 2 and 4

The major advantage of A2C is the performance improvement (reduction in total wait times) achieved with lower training times, as the number of processes was increased. This is due to the parallelism that is incorporated in A2C, as described in Mnih et al. [8].

### 4.5  Summarised Results

A comparison of the parameters as a result of testing the different traffic light agents (RR, MONOPOLY, DQN, A2C) in the three traffic scenarios (SCEN-1, SCEN-2, and SCEN-3) are summarised in Table 2.

**Table 2.** Summarised Results

| Algorithm | Peak Queue Length (vehicles) | | | Fraction of time spent at above half queue length | | |
|---|---|---|---|---|---|---|
| | SCEN-1 | SCEN-2 | SCEN-3 | SCEN-1 | SCEN-2 | SCEN-3 |
| RR | 88 | 56 | 58 | 0.33 | 0.84 | 0.77 |
| MONOPOLY | 89 | 53 | 52 | 0.33 | 0.51 | 0.45 |
| DQN | 90 | 58 | 65 | 0.40 | 0.26 | 0.34 |
| A2C | 84 | 62 | 53 | 0.47 | 0.29 | 0.31 |

### 4.6  Real-World Intersection

The performance of the A2C agent was observed to be the most optimal and scalable and was therefore tested on a simulation of a real-world intersection (JP Nagar Sarakki Signal in Bangalore, Karnataka, India. *Coordinates: 12.906468, 77.573254*). The network was obtained from OpenStreetMaps (OSM) and cleaned up in netedit as shown in Fig. 16. The traffic data was generated as per the Weibull distribution mentioned earlier.

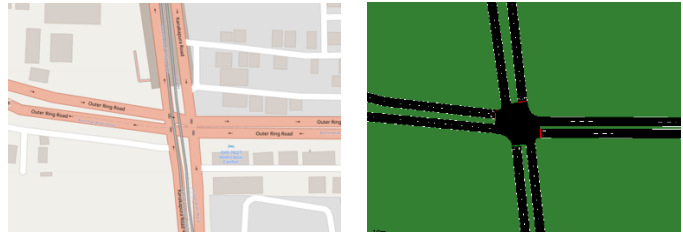

**Fig. 16.** Real-world intersection in OSM (left) and in Netedit (right)

The queue lengths over time were observed for the RR agent (Fig. 17) and the A2C agent (Fig. 18). The peak queue length formed by the A2C agent (35 vehicles) was approximately half of the peak formed by the RR agent (60 vehicles).

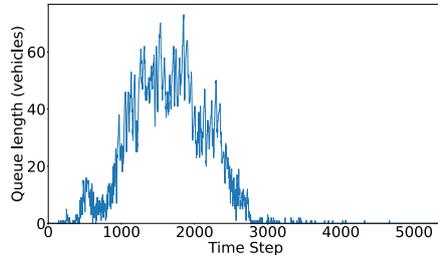
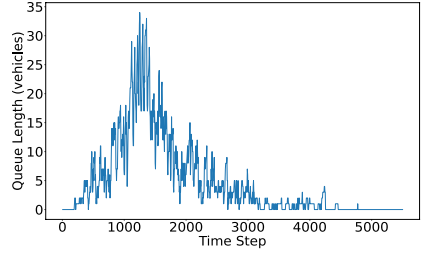

**Fig. 17.** Total queue length vs timestep RR  **Fig. 18.** Total queue length vs timestep A2C

## 5  Conclusion

In this paper, a comprehensive comparative study of some of the techniques and algorithms that can be used for Intelligent Traffic Signal Control (ITSC) has been presented. While Round Robin (RR) is the most prevalent, it is inefficient in clearing out traffic, especially in non-uniform traffic conditions. RR is advantageous as it is computationally simple and can be used in low-density intersections. Feedback Control (MONOPOLY) presents an improvement in non-uniform traffic conditions while staying computationally efficient as well. It can be considered the most efficient when looking at the *performance-computation trade-off*. Deep Q Networks (DQN) is a modern Reinforcement Learning method that yields impressive performance but is computationally intensive. Advantage Actor-Critic (A2C) is an optimization of DQN that performs just as well, while reducing the computation/training time required, by incorporating parallelism in Reinforcement Learning. A2C and other Actor-Critic methods can be best applied in high-density intersections if performance is paramount and computational resources are available.

## References

1. Wey, W. M.: Model Formulation and Solution Algorithm of Traffic Signal Control in an Urban Network. Computers, Environment and Urban Systems. 24(4), 355-378 (2000)
2. Eriskin, E., Karahancer, S., Terzi, S., Saltan, M.: Optimization of Traffic Signal Timing at Oversaturated Intersections Using Elimination Pairing System. In: Procedia Engineering, vol. 187, pp. 295–300 (2017)
3. Abdoos, M., Mozayani, N., Bazzan, A.: Traffic Light Control in Non-Stationary Environments Based on Multi Agent Q-Learning. In: Intelligent Transportation Systems (ITSC), 14th International IEEE Conference (2011)
4. Brys, T., Pham, T., Taylor, M.: Distributed Learning and Multi-Objectivity in Traffic Light Control. Connection Science 26:1, 65-83 (2014)
5. Gregurić, M., Vujić, M., Alexopoulos, C., Miletić, M.: Application of deep reinforcement learning in traffic signal control: An overview and impact of open traffic data. Applied Sciences, 10(11), 4011 (2020)
6. Rasheed, F., Yau, K.L.A., Noor, R.M., Wu, C., Low, Y.C.: Deep Reinforcement Learning for Traffic Signal Control: A Review. IEEE Access (2020)
7. Vidali, A., Crociani, L., Vizzari, G., Bandini, S.: A Deep Reinforcement Learning Approach to Adaptive Traffic Lights Management. In: WOA, pp. 42-50 (2019)
8. Mnih, V., Badia, A., Mirza, M., Graves, A., Lillicrap, T., Harley, T., Silver, D., Kavukcuoglu, K.: Asynchronous Methods for Deep Reinforcement Learning. International Conference on Machine Learning, PMLR (2016)
9. Krajzewicz, D., Hertkorn, G., Rössel, C., Wagner, P.: SUMO (Simulation of Urban MObility)-an open-source traffic simulation. In: Proceedings of the 4th Middle East Symposium on Simulation and Modelling, MESM20002 pp. 183-187 (2002)
10. Alegre, L.: SUMO-RL. GitHub, `github.com/LucasAlegre/sumo-rl` (2019)
11. Wei, H., Zheng, G., Yao, H., Li, Z.: IntelliLight: A Reinforcement Learning Approach for Intelligent Traffic Light Control. In: KDD (2018)
12. Muthupalaniappan, A., Nair, B. S., Rajan, R. A., Krishnan, R. K.: Dynamic Control of Traffic Signals using Traffic Data from Google Maps and Road Cameras. In: International Journal of Recent Technology and Engineering (2019)
13. Zheng, G., Xiong, Y., Zang, X., Feng, J., Wei, H., Zhang, H., Li, Y., Xu, K., Li, Z.: Learning Phase Completion for Traffic Signal Control. In: CKIM, International Conference on Information and Knowledge Management (2019)
14. Hilmani, A., Maizate, A., Hassouni, L.: Automated Real-Time Intelligent Traffic Signal Control System for Smart Cities Using Wireless Sensor Networks. In: Wireless Communications and Mobile Computing (2020)
15. Pandit, K., Ghosal, D., Zhang, H. M., Chuah, C.: Adaptive Traffic Signal Control With Vehicular Ad hoc Networks. In: IEEE Transactions on Vehicular Technology, , vol. 62, no. 4, pp. 1459-1471 (2013).